\documentclass [prb,twocolumn,amsfonts,amssymb,amsmath,superscriptaddress,showpacs]{revtex4}

\usepackage{epsfig}

\newcommand{\mywidth}{0.38\textwidth}
\newcommand{\myheight}{0.28\textwidth}

\begin{document}

\title{Interacting spin waves in the ferromagnetic Kondo lattice model}
\author{A. Schwabe}
\email{andrej.schwabe@physik.uni-hamburg.de}
\affiliation{Festk\"orpertheorie, Institut f\"ur Physik, Humboldt-Universit\"at, 12489 Berlin, Germany}
\affiliation{I. Institut f\"ur Theoretische Physik, Universit\"at Hamburg, 20355 Hamburg, Germany}
\author{W. Nolting}
\affiliation{Festk\"orpertheorie, Institut f\"ur Physik, Humboldt-Universit\"at, 12489 Berlin, Germany}

\begin{abstract}
We present an new approach for the ferromagnetic, three-dimensional, translational-symmetric Kondo lattice model which allows us to derive both magnon energies and linewidths (lifetimes) and to study the properties of the ferromagnetic phase at finite temperatures. Both "anomalous softening" and "anomalous damping" are obtained and discussed.\\
Our method consists of mapping the Kondo lattice model onto an effective Heisenberg model by means of the "modified RKKY interaction" and the "interpolating self-energy approach". The Heisenberg model is approximatively solved by applying the Dyson-Maleev transformation and using the "spectral density approach" with a broadened magnon spectral density.
\end{abstract}


\maketitle

\section{Introduction}

The Kondo lattice model \cite{klm} describes the interaction between two groups of electrons. One group consists of itinerant conduction band electrons which can hop to different lattice sites. The other group concerns localized electrons that couple to a magnetic moment of spin $S$ localized at a certain lattice site. Both subsystems can perform an intra-atomic interaction with each other while neglecting interactions between the itinerant electrons or between the localized spins. For non-degenerated band electrons in real space, the Hamiltonian reads
\begin{subequations}
\begin{align}
	\label{kondo_hamiltonian}
	&H\negthinspace=\negthinspace\negthinspace\sum_{ij\sigma} T_{ij}c^\dagger_{i\sigma}c_{j\sigma}\negthinspace-\negthinspace\frac J2\negthinspace \sum_{i\sigma}\negthinspace\left(z_{\sigma}S^z_i c^\dagger_{i\sigma}c_{i\sigma}\negthinspace+\negthinspace S^{-\sigma}_i c^\dagger_{i\sigma}c_{i-\sigma}\right)\\
	&z_\sigma=\delta_{\sigma\uparrow}-\delta_{\sigma\downarrow},\;S^\sigma=S^+\delta_{\sigma\uparrow}+S^-\delta_{\sigma\downarrow}
\end{align}
\end{subequations}
where $c^{(\dagger)}_{i\sigma}$ represents an annihilition (creation) operator for an electron of spin projection $\sigma$ at a lattice site $\mathbf R_i$. $J$ is the Hund's coupling constant and $T_{ij}$ are the hopping integrals. Since we are investigating the ferromagnetic Kondo lattice model, $J>0$.\\
The second term in Eq. \eqref{kondo_hamiltonian} describes an Ising-like interaction between the $z$-components of the localized and the itinerant spin. The third term accounts for the spin exchange processes between the two subsystems.\\
We will treat the three-dimensional, translational-symmetric, infinitely-extended Kondo lattice model.\\
\\
The Kondo lattice model is believed to characterize the basic physics of a wide variety of solid state materials.\\
Magnetic semiconductors, e.g., EuO, are a prominent class of substances, which draw notable attention due to the "red shift" of the optical absorption edge upon cooling from $T=T_\text C$ to $T=0K$. One can conclude that the coupling constant $J$ is positive and of the order of some tenth of $eV$. In contrast, the magnetic ordering of the localized spins is explained via special superexchange mechanisms.\\
Ferromagnetic local moment metals, such as Gd, are another application. An RKKY-(Ruderman and Kittel\cite{rkky3}, Kasuya\cite{rkky1}, Yosida\cite{rkky2}) type interaction is supposed to create the ferromagnetic order. The magnetism relies on localized $4f$ electrons that are shielded from the $4f$ orbitals of adjacent atoms by other completely filled orbitals. On the other side, the conductivity properties are determined by itinerant $5d$ or $6s$ electrons.\\
The discovery of the "colossal magnetoresistance" (CMR) and its promising technological application motivated a considerable research effort that is related to the manganese oxides with perovskite structures $T_{1-x}D_x$MnO$_3$ (T=La, Pr, Nd; D=Sr, Ca, Ba, Pb). A prototype is the well-known compound La$_{1-x}$Ca$_x$MnO$_3$ which can be obtained by replacing a trivalent La$^{3+}$ ion with the divalent earth-alkali ion Ca$^{2+}$ in La$^{3+}$Mn$^{3+}$O$_3$ leading to a homogeneous valence mixture of the manganese ions Mn$^{3+}_{1-x}$Mn$^{4+}_x$. The three $3d-t_{2g}$ electrons of Mn$^{4+}$ are considered as localized forming a spin of $S=\frac32$. They are coupled to the $n=(1-x)$ itinerant $3d-e_g$ electrons per Mn site by a ferromagnetic coupling $J>0$. $J$ is estimated to be much larger than the electronic bandwidth since the manganites are bad electrical conductors.\\
\\
Many fascinating features of the Kondo lattice model can be accredited to the complex correlation between the magnetic and electronic subsystem. In this regard, one challenging issue represents the "anomalous softening" of the spin wave dispersion, which has attracted comprehensive interest. The spin wave dispersion relation of manganites with high $T_\text C$ resembles one of a simple Heisenberg model with nearest-neighbour exchange only.\cite{furukawa,zhang_review} However, some manganites with lower $T_\text C$ exhibit apparent deviations from this behaviour, that are strongly dependent on the band occupation.\cite{zhang_review,dai,ye2,moussa} Despite extensive theoretical work in this field, the softening of the dispersion relation near the boundaries of the Brillouin zone still lacks a complete explanation. Currently, disorder induced softening has been excluded for some materials.\cite{ye2, zhang_review} On the other hand, the incorporation of an antiferromagnetic super exchange interaction between the Mn ions into the Hamiltonian of the Kondo lattice model has been proposed to take into account the antiferromagnetic tendencies of the parent material LaMnO$_3$.\cite{mancini}\\
In recent years, unusually large magnon damping at the Brillouin zone boundaries and low temperatures has come to the fore. This is frequently referred to as "anomalous damping."\cite{zhang_review, dai} Evidence has been found in neutron scattering experiments with manganites and raised questions concerning the nature of anomalous damping and its link to anomalous softening. Besides the electron-magnon interaction, some authors speculate about a magnon-phonon coupling for certain manganites as an origin,\cite{dai} while other authors reject it.\cite{moussa} Thus, it is of crucial importance to develop new spin wave theories for the Kondo lattice model and to study whether anomalous damping can be traced back to the electron-magnon interaction.\\
\\
In this work, we concentrate on the magnetic subsystem of the Kondo lattice model. The aim is to investigate the dependencies of the energy as well as the linewidths of the elementary magnetic excitations called spin waves or magnons, respectively. A subsection of the paper will treat the anomalous features of the magnon spectrum mentioned above and include a discussion of the influence of temperature.\\
We will introduce a new solution for the Kondo lattice model which is as well applicable to the Heisenberg model. It goes explicitly beyond standard methods like the "random phase approximation," by accounting for correlations of higher order. Although non-pertubative, it is still controlled in the sense that, in principle, it results from the moments of an exact high energy expansion. We assume quantum spins, so our theory is not restricted to the classical limit of large spin values $S\gg1$.\\
\\
The paper is structured as follows: First, we will demonstrate how the Kondo lattice model is mapped onto a Heisenberg model (Sec. \ref{map}). Both employed theories, the "modified RKKY interaction" and the "interpolating self-energy approach", have been already successfully applied to the Kondo lattice model for various other problems.\cite{santos, stier1, nol_mrkky_mcda, sandschneider, kreissl} They will fix the electron-spin interaction. In Sec. \ref{sol}, we will focus on the spin-spin interaction by introducing a new solution for the Heisenberg model that will not only allow us to calculate the energy of the magnetic excitations, but also their linewidth. Sec. \ref{kondo} will proceed with numerical results for the Kondo lattice model that provide insights into the properties of its ferromagnetic phase and an investigation of the dependence on the intra-atomic coupling constant $J$, the conduction band occupation $n$, and the temperature $T$.

\section{\label{map}Mapping onto a Heisenberg model}

The Kondo lattice model provokes a complex many-body problem solvable only in a few limiting cases. Hence, we try mapping the Hamiltonian of the Kondo lattice model \eqref{kondo_hamiltonian} onto an effective Heisenberg Hamiltonian in which the conduction band electrons mediate the indirect exchange interaction between the localized spins. The idea is to use the "modified RKKY interaction\cite{henning_mrkky,nol_mrkky_mcda}" (mRKKY), wherein the Hamiltonian is averaged in the electronic subspace
\begin{align}
\label{H_sf}
	H_\text{s}=-\frac J2\sum_{i\sigma}\left(z_\sigma S^z_i\langle c^\dagger_{i\sigma}c_{i\sigma}\rangle^\text{(s)}+S_i^{-\sigma}\langle c^\dagger_{i\sigma}c_{i-\sigma}\rangle^\text{(s)}\right).
\end{align}
The arising expectation value $\langle c_{i\sigma}^\dagger c_{i-\sigma}\rangle^\text{(s)}$ does not vanish generally since the spin conservation is valid for the total system of the localized spins and the itinerant electrons while the averaging is done in the electronic subspace only. The expectation values can be calculated by using the spectral theorem and the corresponding electron Green's functions called "restricted Green's functions"
\begin{subequations}
\begin{align}
\langle c^\dagger_{i\sigma}c_{i\sigma}\rangle^\text{(s)}&=-\frac{1}{\pi\hbar}\int dE\ f_-(E)\ \text{Im}\ G^{\sigma\sigma,\text{(s)}}_{ii}\\
G^{\sigma\sigma,\text{(s)}}_{ij}&=\langle\langle c_{i\sigma};c^\dagger_{j\sigma}\rangle\rangle^{\text{(s)}}
\end{align}
\end{subequations}
\begin{subequations}
\begin{align}
\langle c^\dagger_{i\sigma}c_{i-\sigma}\rangle^\text{(s)}&=-\frac{1}{\pi\hbar}\int dE\ f_-(E)\ \text{Im}\ G^{-\sigma\sigma,\text{(s)}}_{ii}\\
G^{-\sigma\sigma,\text{(s)}}_{ij}&=\langle\langle c_{i-\sigma};c^\dagger_{j\sigma}\rangle\rangle^{\text{(s)}},
\end{align}
\end{subequations}
where $f_-(E)$ is the Fermi function. After introducing the free Green's function for non-interacting electrons $G_{ij}^{(0)}(E)$, the equations of motion of $G_{ij}^{\sigma\sigma,\text{(s)}}(E)$ and $G_{ij}^{-\sigma\sigma,\text{(s)}}(E)$ can be solved
\begin{align}
\label{rGF1}
G_{ij}^{\sigma\sigma,\text{(s)}}&(E)=G^{(0)}_{ij}(E)-\frac J2\sum_lG_{il}^{(0)}\cdot\\\nonumber
	&\cdot\left(z_\sigma S^z_l\langle\langle c_{l\sigma};c^\dagger_{j\sigma}\rangle\rangle^\text{(s)}+S_l^{-\sigma}\langle\langle c_{l-\sigma};c^\dagger_{j\sigma}\rangle\rangle^\text{(s)}\right)\\
\label{rGF2}
G_{ij}^{-\sigma\sigma,\text{(s)}}&(E)=-\frac J2\sum_lG_{il}^{(0)}\cdot\\\nonumber
	&\cdot\left(-z_{\sigma}S^z_l\langle\langle c_{l-\sigma};c^\dagger_{j\sigma}\rangle\rangle^\text{(s)}+S_l^{\sigma}\langle\langle c_{l\sigma};c^\dagger_{j\sigma}\rangle\rangle^\text{(s)}\right).
\end{align}
Now we replace the restricted Green's functions on the right-hand sides of Eqs. \eqref{rGF1} and \eqref{rGF2} with their full equivalents
\begin{align}
&G_{ij}^{\sigma\sigma,\text{(s)}}(E)\rightarrow G_{ij\sigma}(E)=\langle\langle c_{i\sigma};c^\dagger_{j\sigma}\rangle\rangle\\
&G^{-\sigma\sigma,\text{(s)}}_{ij}(E)\rightarrow G^{-\sigma\sigma}_{ij}(E)=0.
\end{align}
$G^{-\sigma\sigma}_{ij}(E)$ vanishes because of spin conservation. $G_{ij\sigma}(E)$ labels the Green's function of interacting electrons.\\
These solutions are inserted into the averaged Hamiltonian \eqref{H_sf}. Eventually, our approach leads to an effective Heisenberg Hamiltonian\footnote{We use the identity $\varepsilon_\mathbf q=\varepsilon_{-\mathbf q}$ that is valid for the cubic lattices.} 
\begin{align}
\label{eff_Hamiltonian}
	H_{\text{Kondo}}\;\stackrel{\text{mRKKY}}{\longrightarrow}\;H_\text{eff}=-\sum_{ij} J_{ij}\mathbf S_i\cdot \mathbf S_j.
\end{align}
The effective exchange integrals are functionals of the electronic self-energy $\Sigma_{ij\sigma}(E)$ via the electron Green's function $G_{ij\sigma}(E)$
\begin{align}
\label{exchange_integrals}
		J_{ij}=\frac{J^2}{4\pi\hbar^2}\int dE\ f_-(E)\ \text{Im}\ \sum_\sigma G_{ij}^{(0)}(E)G_{ij\sigma}(E).
\end{align}
The replacements \eqref{rGF1} and \eqref{rGF2} comprise many-body correlations of higher order than the conventional RKKY theory that would be obtained by replacing
\begin{align}
G_{ij}^{\sigma\sigma,\text{(s)}}(E)\rightarrow G^{(0)}_{ij}(E),\;\;\;\;\;G^{-\sigma\sigma,\text{(s)}}_{ij}(E)\rightarrow 0.
\end{align}
\\
We use the interpolating self-energy approach\cite{nol_isa1} (ISA) in order to set the electronic self-energy $\Sigma_{ij\sigma}(E)$. It is derived for the limiting cases of the ferromagnetically-ordered semiconductor, the atomic limit, and second order pertubation theory assuming vanishing band occupation $n=\sum_\sigma\langle c^\dagger_{i\sigma}c_{i\sigma}\rangle\rightarrow0$. An interpolation between the limiting cases performed by fitting leading terms in its rigorous high-energy expansion provides the result
\begin{subequations}
\begin{align}
\label{self_energy}
		\Sigma_{ij\sigma}(E)=&-\frac J2z_\sigma\langle S^z\rangle\delta_{ij}+\\\nonumber
		&+\frac{J^2}{4}\frac{a_\sigma G^{(0)}_{ii}\left(E-\frac12Jz_\sigma\langle S^z\rangle\right)}{1-b_\sigma G^{(0)}_{ii}\left(E-\frac12Jz_\sigma\langle S^z\rangle\right)}\delta_{ij}\\
		a_\sigma=S(S+1)&-z_\sigma\langle S^z\rangle(z_\sigma\langle S^z\rangle+1),\;b_\sigma=\frac J2.
\end{align}
\end{subequations}
Although derived in the low concentration limit, we apply the self-energy \eqref{self_energy} to the case of finite band occupations $n>0$.\\\\
In summary, the problem is reduced to the solution of an effective Heisenberg model with exchange integrals $J_{ij}$ that depend on the coupling constant $J$, the band occupation $n$, and the temperature $T$: $J_{ij}=J_{ij}(J,n,T)$.

\section{\label{sol}The effective Heisenberg model}
\subsection{Solution}

It is convenient to transform the spin operators $\mathbf S_i$ of the effective Heisenberg Hamiltonian \eqref{eff_Hamiltonian} into bosonic magnon operators $\left\lbrace a_i,a^\dagger_i,n_i=a^\dagger_ia_i\right\rbrace$ by means of the Dyson-Maleev transformation \cite{dyson1,dyson2,maleev}:
\begin{subequations}
\begin{align}
\label{dmtrafo1}
	S^+_i&=\sqrt{2S}a_i,\;S^-_i=\sqrt{2S}a^\dagger_i\left(1-\frac{n_i}{2S}\right)\\
\label{dmtrafo2}
	S^z_i&=S-n_i.
\end{align}
\end{subequations}
After a Fourier transformation, the Heisenberg Hamiltonian then reads in momentum space
\begin{align}
\label{DM}
&H_\text{Heisenberg}\;\longrightarrow\;H_\text{DM}=\frac1N\sum_{\mathbf{q}^\prime}\hbar\omega_{\mathbf{q}^\prime}n_{\mathbf{q}^\prime}+\\\nonumber
&+\frac{1}{N^4}\sum_{\mathbf{q}_1\ldots\mathbf{q}_4}(J_{\mathbf{q}_4}-J_{\mathbf{q}_1-\mathbf{q}_3})a^\dagger_{\mathbf{q}_1}a^\dagger_{\mathbf{q}_2}a_{\mathbf{q}_3}a_{\mathbf{q}_4}\delta_{\mathbf{q}_1+\mathbf{q}_2,\mathbf{q}_3+\mathbf{q}_4}
\end{align}
where $\hbar\omega_\mathbf q=2S(J_0-J_\mathbf q)$ stands for the bare energy of a free magnon with momentum $\mathbf q$ and $N$ for the number of lattice sites. The second summand of $H_\text{DM}$ in Eq. \eqref{DM} describes the magnon-magnon interaction and causes the existence of finite linewidths and the renormalization of the magnon energy. The Dyson-Maleev transformation makes it possible to take the complete interaction between the magnons into account without making approximations that are necessary for other theories, e.g., the Holstein-Primakoff transformation\cite{holstein_primakoff}.\\
At this stage, we need to mention that the transformation \eqref{dmtrafo1} and \eqref{dmtrafo2} leads to unphysical states for temperatures near the transition temperature $T_\text C$ since we transform from a Hilbert space which is $(2S+1)$ dimensional into one with infinite dimensions. Nevertheless, according to Dyson, the contributions to the free energy from these unphysical states are smaller than $\exp\left(-\alpha \frac{T_\text C}{T}\right)$ where $\alpha$ is a coefficient of order unity.\cite{dyson2}\\
Additionally, $S^+_i$ and $S^-_i$ are not Hermitian conjugated in the Dyson-Maleev formalism. However, Bar'yakhtar et al.\cite{bar'yakhtar} showed that the contributions to spin correlation functions from unphysical states arising from the non-Hermiticity are of the order $\exp\left(-\frac{T^*}{T}\right)$ where $k_\text BT^*=S(2S+1)J_0$.\\
\\
The bosonic Heisenberg model \eqref{DM} is solved by applying the "spectral density approach." The spectral moments of the spectral density $S_\mathbf q(E)$ are defined by
\begin{align}
\label{moments}
M^{(n)}_\mathbf{q} =\frac1\hbar\int dE\ E^n S_\mathbf{q}(E),
\end{align}
but they can also be evaluated exactly and independently from Eq. \eqref{moments} by the following relation
\begin{align}
\label{sda}
M^{(n)}_\mathbf{q}&=\\\nonumber
&=\langle [ \underbrace{\left[\left[a_\mathbf{q},H\right]_-,\ldots, H\right]_-}_{p\text{-fold commutator}},\underbrace {\left[H,\ldots,\left[H,a^\dagger_\mathbf{q}\right]_-\right]_-}_{(n-p)\text{-fold cmmutator}} ]_{-} \rangle.
\end {align}
The approach requires a spectral density which is usually guessed, e.g., from experiments or theoretical considerations. Parameters of $S_\mathbf q(E)$ can be evaluated by a sufficient set of equations that is derived from the equivalence of Eqs. \eqref{moments} and \eqref{sda}.\\
In our case, $S_\mathbf q(E)$ represents the magnon spectral density which is associated with the average magnon occupation number by the spectral theorem
\begin{align}
	\langle a^\dagger_\mathbf qa_\mathbf q\rangle=\langle n_\mathbf{q}\rangle=\frac1\hbar\int dE\ f_+(E) S_\mathbf{q}(E),
\end{align}
where $f_+(E)$ is the Bose function. Since we are interested in lifetime effects, we need to fit the first three spectral moments $M_\mathbf{q}^{(n\leq2)}$ and use a spectral density of finite width. The renormalized magnon energies $\hbar\Omega_\mathbf q$ and their spectral linewidths $\Gamma_\mathbf q$ or lifetimes, respectively,
\begin{align}
	\tau_\mathbf q=\frac{\hbar}{\Gamma_\mathbf q}
\end{align}
work as parameters that need to be calculated from the set of equations.\\
For simplicity and without loss of generality, we want to restrict our derivation to a symmetric spectral density $S_\mathbf q (\hbar\Omega_\mathbf q+E)=S_\mathbf q (\hbar\Omega_\mathbf q-E)$.\footnote{The restriction to a symmetric spectral density $S_\mathbf q (\hbar\Omega_\mathbf q+E)=S_\mathbf q (\hbar\Omega_\mathbf q-E)$ facilitates the transition from the spectral moments $M_\mathbf q$ to $\hbar\Omega_\mathbf q$ and $\Gamma_\mathbf q$. The corresponding substitution in the integral in \eqref{moments} is $\frac{E-\hbar\Omega_\mathbf q}{\Gamma_\mathbf q}\rightarrow E$} This is in agreement with neutron scattering experiments\cite{exp3} and other theories,\cite{tahir} where $S_\mathbf q(E)$ has the approximate shape of a Lorentzian
\begin{align}
\label{Lorentzian}
	S^{\text{Lorentzian}}_\mathbf{q}(E)=\frac{\hbar\Gamma_\mathbf q}{\pi}\frac{1}{(E-\hbar\Omega_\mathbf q)^2+\Gamma_\mathbf q^2}
\end{align}
or a Gaussian, respectively,
\begin{align}
\label{Gaussian}
	S^{\text{Gaussian}}_\mathbf{q}(E)=\frac{\hbar}{\sqrt{\pi}\Gamma_\mathbf{q}}e^{-\left(\frac{E-\hbar\Omega_\mathbf{q}}{\Gamma_\mathbf{q}}\right)^2}.
\end{align}
One must keep in mind that the Lorentzian must be restricted to a finite energy interval ensuring a finite second spectral moment $M^{(2)}_\mathbf q$.\\
The zeroth spectral moment
\begin{align}
	M_\mathbf{q}^{(0)}&=\langle \left[a_\mathbf{q},a^\dagger_\mathbf{q}\right]_-\rangle=1
\end{align} 
expects a normalized spectral density $S_\mathbf q(E)$ according to Eq. \eqref{moments}. For the first spectral moment we get:
\begin{align}
\label{M1}
	M_\mathbf{q}^{(1)}&=\langle \left[\left[ a_\mathbf{q},H_\text{DM} \right]_-,a^\dagger_\mathbf{q}\right]_-\rangle\\\nonumber
	&=\hbar\omega_\mathbf{q}M^{(0)}_\mathbf{q}+\frac{2}{N}\sum_{\mathbf{q}^\prime}(J_\mathbf{q}+J_{\mathbf{q}^\prime}-J_0-J_{\mathbf{q}^\prime-\mathbf{q}})\langle n_{\mathbf{q}^\prime}\rangle.
\end{align}
The result for the second spectral moment is
\begin{align}
\label{preM2}
M_\mathbf{q}^{(2)}=&\langle \left[\left[ a_\mathbf{q},H_\text{DM} \right]_-,\left[ H_\text{DM},a^\dagger_\mathbf{q}\right]_-\right]_-\rangle\\\nonumber
	=&2\hbar\omega_\mathbf{q} M^{(1)}_\mathbf{q}-\left(\hbar\omega_\mathbf{q}M^{(0)}_\mathbf q\right)^2+\frac{1}{N^4}\cdot\\\nonumber
&\cdot\sum_{\mathbf{q}_1\ldots\mathbf{q}_4}(J_{\mathbf q_3}+J_{\mathbf q_4}-J_{\mathbf q_3-\mathbf q}-J_{\mathbf q_4-\mathbf q})\cdot\\\nonumber
	&\;\;\;\;\cdot(J_\mathbf q+J_{\mathbf q_1+\mathbf q_2-\mathbf q}-J_{\mathbf q_1-\mathbf q}-J_{\mathbf q_2-\mathbf q})\cdot\\\nonumber
	&\;\;\;\;\cdot\Bigg( 2\langle a^\dagger_{\mathbf{q}_3+\mathbf{q}_4-\mathbf{q}}a_{\mathbf{q}_3}a^\dagger_{\mathbf{q}_2}a_{\mathbf{q}_1+\mathbf{q}_2-\mathbf{q}}\rangle\delta_{\mathbf{q}_1,\mathbf{q}_4}+\\
\nonumber
&\;\;\;\;\;\;\;\;+2\langle a^\dagger_{\mathbf{q}_3+\mathbf{q}_4-\mathbf{q}}a^\dagger_{\mathbf{q}_2}a_{\mathbf{q}_3}a_{\mathbf{q}_1+\mathbf{q}_2-\mathbf{q}}\rangle\delta_{\mathbf{q}_1,\mathbf{q}_4}-\\\nonumber
	&\;\;\;\;\;\;\;\;-\langle a^\dagger_{\mathbf{q}_1}a^\dagger_{\mathbf{q}_2}a_{\mathbf{q}_3}a_{\mathbf{q}_4}\rangle\delta_{\mathbf{q}_1+\mathbf{q}_2,\mathbf{q}_3+\mathbf{q}_4}\Bigg).
\end {align}
A simple ansatz for the unknown, higher expectation values in Eq. \eqref{preM2}, such as $\langle a^\dagger_{\mathbf{q}_3+\mathbf{q}_4-\mathbf{q}}a_{\mathbf{q}_3}a^\dagger_{\mathbf{q}_2}a_{\mathbf{q}_1+\mathbf{q}_2-\mathbf{q}}\rangle$, is derived by decoupling them using a mean field approximation with respect to momentum conservation
\begin{align}
	\langle &a^\dagger_{\mathbf{q}_3+\mathbf{q}_4-\mathbf{q}}a_{\mathbf{q}_3}a^\dagger_{\mathbf{q}_2}a_{\mathbf{q}_1+\mathbf{q}_2-\mathbf{q}}\rangle\delta_{\mathbf{q}_1,\mathbf{q}_4}\\\nonumber
				&\stackrel{MF}{\longrightarrow}\langle a^\dagger_{\mathbf{q}_3+\mathbf{q}_4-\mathbf{q}}a_{\mathbf{q}_3}\rangle\langle a^\dagger_{\mathbf{q}_2}a_{\mathbf{q}_1+\mathbf{q}_2-\mathbf{q}}\rangle\delta_{\mathbf{q}_1,\mathbf{q}_4}\delta_{\mathbf{q}_4,\mathbf{q}}+\\\nonumber
				&\thickspace\thickspace\thickspace\thickspace\thickspace+\langle a^\dagger_{\mathbf{q}_3+\mathbf{q}_4-\mathbf{q}}a_{\mathbf{q}_1+\mathbf{q}_2-\mathbf{q}}\rangle\langle a^\dagger_{\mathbf{q}_2}a_{\mathbf{q}_3}\rangle\delta_{\mathbf{q}_1,\mathbf{q}_4}\delta_{\mathbf{q}_2,\mathbf{q}_3}\\\nonumber
&\ldots
\end{align}
Therewith, the solution is formally completed:
\begin{align}
\label{M2}
M_\mathbf{q}^{(2)}=&\left(\hbar\omega_\mathbf{q}M^{(0)}_\mathbf q\right)^2+\frac{1}{N^2}\cdot\\\nonumber
&\cdot\sum_{\mathbf q^\prime}\sum_{\mathbf q^{\prime\prime}}\Big(2(J_{\mathbf{q}^\prime}+J_{\mathbf{q}^{\prime\prime}-\mathbf{q}^\prime+\mathbf{q}}-J_{\mathbf{q}^{\prime}-\mathbf{q}}-J_{\mathbf{q}^{\prime\prime}-\mathbf{q}^\prime})\cdot\\\nonumber
&\;\;\cdot(J_\mathbf{q}+J_{\mathbf{q}^{\prime\prime}}-J_{\mathbf{q}^{\prime}-\mathbf{q}}-J_{\mathbf{q}^{\prime\prime}-\mathbf{q}^\prime})-\\\nonumber
&\;\;-(J_{\mathbf{q}^\prime}+J_{\mathbf{q}^{\prime\prime}}-J_{\mathbf{q}^{\prime}-\mathbf{q}}-J_{\mathbf{q}^{\prime\prime}-\mathbf{q}})\cdot\\\nonumber
&\;\;\cdot(J_\mathbf{q}+J_{\mathbf{q}^\prime+\mathbf{q}^{\prime\prime}-\mathbf{q}}-J_{\mathbf{q}^{\prime}-\mathbf{q}}-J_{\mathbf{q}^{\prime\prime}-\mathbf{q}})\Big)\langle n_{\mathbf{q}^\prime}\rangle\langle n_{\mathbf{q}^{\prime\prime}}\rangle.
\end{align}
For numerical reasons, we still need to simplify Eq. \eqref{M2} to prevent eight-dimensional integrals in expressions like $\frac{1}{N^2}\sum_{\mathbf q^\prime}\sum_{\mathbf q^{\prime\prime}}J_{\mathbf{q}^{\prime\prime}-\mathbf{q}^\prime+\mathbf{q}}J_{\mathbf{q}^{\prime\prime}}\langle n_{\mathbf{q}^\prime}\rangle\langle n_{\mathbf{q}^{\prime\prime}}\rangle$. This is done by exploiting the translational symmetry:
\begin{subequations}
\begin{align}
	J_0-J_\mathbf q&=\negthickspace\sum_{\text{shells }i}\negthickspace z_iJ_i\left(1-\gamma^{(i)}_\mathbf q\right)\\
	\gamma^{(i)}_\mathbf{q}&=\frac{1}{z_i}\sum_{\mathbf R\text{ of}\atop\text{shell }i}e^{\text i\mathbf{q}\cdot\mathbf R}.
\end{align}
\end{subequations}
A shell is defined by all lattice sites $\mathbf R$ at the same distance $|\mathbf R-\tilde{\mathbf R}|$ to an offset lattice site $\tilde{\mathbf R}$.\footnote{The lattice sites can be ordered by any other pattern as well. Anyhow, for an isotropic magnetic system with $J_{ij}=J_{ij}(|\mathbf R_i-\mathbf R_j|)$ it seems natural to require $|\mathbf R-\tilde{\mathbf R}|=\text{const}$ for a shell.} Thereby $z_i$ denotes the number of all lattice sites of shell $i$ and $J_i$ the exchange integral of shell $i$. The shells are numbered and sorted by the size of their radii, i.e. $i=1$ stands for the nearest neighbours, $i=2$ for the next nearest neighbours etc. Within the shell concept, the problematic terms easily factorize\footnote{We use the identity $\gamma_\mathbf q=\gamma_{-\mathbf q}$ that is valid for the cubic lattices.}
\begin{align}
&\frac{1}{N^2}\sum_{\mathbf{q}^\prime\mathbf{q}^{\prime\prime}}\gamma^{(i)}_{\mathbf{q}^{\prime\prime}-\mathbf{q}^{\prime}-\mathbf{q}_1}\gamma^{(j)}_{\mathbf{q}^{\prime\prime}-\mathbf{q}_2}\langle n_{\mathbf{q}^\prime}\rangle\langle n_{\mathbf{q}^{\prime\prime}}\rangle\\\nonumber
&=\gamma^{(i)}_{\mathbf{q}_1}\gamma^{(j)}_{\mathbf{q}_2}\frac1N\sum_{\mathbf q^{\prime}}\left(\gamma^{(i)}_{\mathbf{q}^{\prime}}\langle n_{\mathbf{q}^{\prime}}\rangle\right)\negthickspace\negthickspace \sum_{\text{shells } m}\negthickspace\negthickspace N_{ijm}\frac1N\sum_{\mathbf{q}^{\prime\prime}}\gamma^{(m)}_{\mathbf{q}^{\prime\prime}}\langle n_{\mathbf{q}^{\prime\prime}}\rangle
\end{align}
where we use a notation similar to Dvey-Aharon and Fibich \cite{fibich}:
\begin{align}
N_{ijm}=\frac{1}{z_iz_j}\sum_{\mathbf R_i,\mathbf R_j}\negthickspace\delta_{\mathbf R_i+\mathbf R_j,\mathbf R_m}.
\end{align}
$N_{ijm}$ gives the number of shell vectors $\mathbf R_m$ that can be constructed out of the sum of shell vectors of the shells $i$ and $j$
\begin{align}
\mathbf R_i+\mathbf R_j=\mathbf R_m.
\end{align}
By applying the shell concept not only to the expressions of the second spectral moment $M_\mathbf q^{(2)}$ in Eq. \eqref{M2}, but also to the first spectral moment $M_\mathbf q^{(1)}$ in Eq. \eqref{M1}, we finally obtain after some algebra
\begin{subequations}
\begin{align}
\label{results1}
	\hbar\Omega_\mathbf{q}=&2\negthickspace\negthickspace\sum_{\text{shells }i}\negthickspace\negthickspace 				z_iJ_i\left(1-\gamma_\mathbf{q}^{(i)}\right)(S-A_i)\\
\label{results2}
	\Gamma^2_\mathbf{q}\tilde m_\mathbf q= &4\negthickspace\negthickspace\sum_{\text{shells }i,j}\negthickspace\negthickspace z_iz_jJ_iJ_jF_{ij}\cdot\\\nonumber
&\cdot\left(1-\gamma^{(i)}_\mathbf q\right)\left(F_{ij}+(A_j-A_i)\left(1-\gamma_\mathbf q^{(j)}\right)\right).
\end{align}
\end{subequations}
The influence of the shape of $S_\mathbf q(E)$ on the second spectral moment $M_\mathbf q^{(2)}$ is contained in the dimensionless quantity $\tilde m_\mathbf q$:
\begin{align}
\tilde m_\mathbf q=\frac1\hbar\int dx\ x^2\Gamma_\mathbf qS_\mathbf q(x\Gamma_\mathbf q+\hbar\Omega_\mathbf q).
\end{align}
When using a Lorentzian or a Gaussian spectral density, $\tilde m_\mathbf q$ is advantageously independent on the momentum $\mathbf q$.\\
$F_{ij}$ is merely determined by the $A_i$:
\begin{align}
F_{ij}=A_0-A_i-N_{ij0}A_0-\negthickspace\negthickspace\negthickspace\negthickspace\sum_{\text{shells }m>0}\negthickspace\negthickspace\negthickspace\negthickspace N_{ijm}(A_0-A_m).
\end{align}
Therewith, the $A_i$ remain the only unknown quantities in Eq. \eqref{results1} and \eqref{results2} at a given temperature. Because of the definition
\begin{align}
\label{Ai_def}
A_i&=\begin{cases}
     	\frac1N\sum_{\mathbf q^\prime}\langle n_{\mathbf q^\prime}\rangle	& \text{if }i=0\\
	\frac1N\sum_{\mathbf q^\prime}\left(1-\gamma^{(i)}_{\mathbf q^\prime}\right)\langle n_{\mathbf q^\prime}\rangle	&\text{if }i>0\\
     \end{cases}
\end {align}
the $A_i$ depend on the spectral density $S_\mathbf q(E)$ and via the Eqs. \eqref{results1} and \eqref{results2} on each other 
\begin{align}
\label{self-consistently}
A_i=A_i(\left\lbrace A_j\right\rbrace).
\end{align}
Therefore, they have to be self-consistently calculated. With the solution satisfying Eq. \eqref{self-consistently}, one can directly compute $\hbar\Omega_\mathbf q$ and $\Gamma_\mathbf q$ for a given momentum $\mathbf q$.

\begin{figure}[t]
	\centering
\includegraphics[width=\mywidth,height=\myheight]{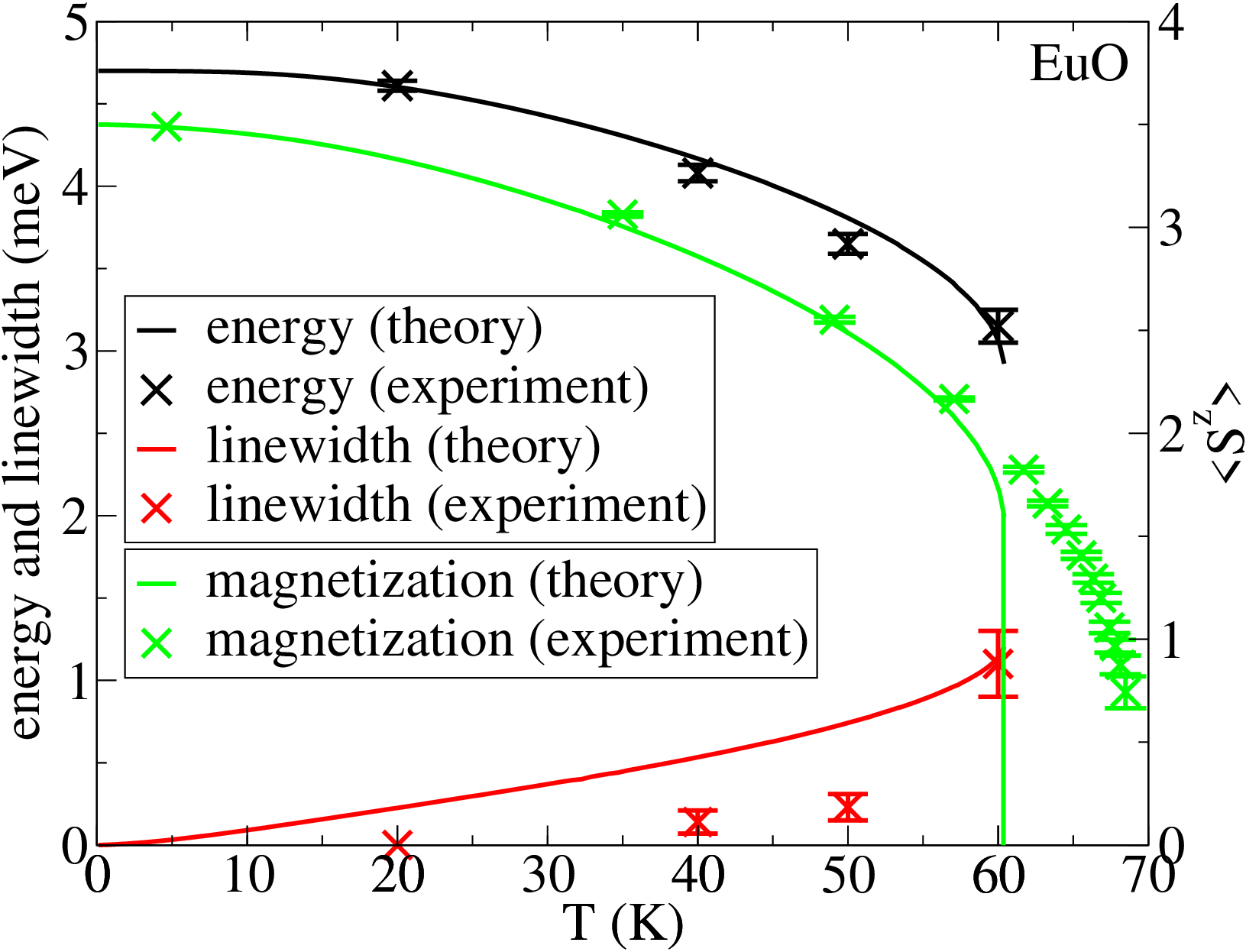}
\caption{(Color online) Temperature dependence of the magnon energy $\hbar\Omega_q(T)$ and the linewidth $\Gamma_q(T)$ for $q=|\mathbf q|=0.8\text \AA^{-1}$, and of the magnetization $\langle S^z\rangle(T)$. The calculations (lines) are carried out for the parameters of EuO (ref. \onlinecite{exp1}) in comparison to experimental data from the Refs. \onlinecite{exp3}, \onlinecite{glinka}, and \onlinecite{exp2} (crosses). As in the experimental analysis, a Lorentzian \eqref{Lorentzian} is used for the magnon spectral density $S_\mathbf q(E)$.}
\label{comparison}
\end{figure}

\subsection{\label{compare_EuO}Comparison to experimental data and other theories}

In terms of checking our theory for a real system, we consider the Heisenberg ferromagnet EuO, whose exchange integrals $J_1$ and $J_2$ are known.\cite{exp1} According to Fig. \ref{comparison}, a good agreement concerning the magnon properties $\hbar\Omega_\mathbf q$ and $\Gamma_\mathbf q$, and the magnetization $\langle S^z\rangle$ can be found between the numerical results of our theory and the experimental data for a wide range of low and intermediate temperatures and for not-too-small momenta.\cite{exp1,exp2,exp3,glinka} At temperatures near $T_\text C$, the unphysical states cause a wrong first order phase transition that contradicts the experimental data.\\
Results similar to Eq. \eqref{results2} have been obtained by other theories of the Heisenberg model,\cite{cooke,marshall,harris,tahir} but one notes differences for a small range of small momenta $\mathbf q\rightarrow\mathbf 0$. There, the linewidths of our formula \eqref{results2} turn out to be too large: $\Gamma^2_\mathbf q\stackrel{\mathbf q\rightarrow\mathbf 0}{\sim}\mathbf q^2$ - other authors\cite{cooke} propose at least a dependence $\Gamma_\mathbf q\stackrel{\mathbf q\rightarrow\mathbf 0}{\sim}\mathbf q^2$. This discrepancy must be classified as a consequence of our approximations. Furthermore, our results give $\Gamma_X(T)\sim T^{1.4}$ while the authors of Refs. \onlinecite{cooke} and \onlinecite{marshall} suggest a stronger dependence $\Gamma_\mathbf q\sim T^3$.

\begin{table}[b]
\centering
\begin{tabular}{|l|c|}	\hline
	lattice structure&simple cubic (sc)\\
	spin value $S$&$\frac32$\\
	magnon spectral density $S_\mathbf q(E)$&Gaussian \eqref{Gaussian}\\
	conduction band&$s$ band, bandwidth $W=1eV$,\\
	&tight-binding\\\hline
\end{tabular}
\caption{Setting of the main parameters used in the calculations for Sec. \ref{kondo}.}
\label{settings}
\end{table}

\section{\label{kondo}The Kondo lattice model}

In order to circumvent the problem of too many possible parameter combinations to discuss, we have chosen three exemplary regions. Small band occupations should be a suitable criterion for ferromagnetic semiconductors and $J=W$ for manganites, where $W$ is the electronic bandwidth. Intermediate $J$ and intermediate $n$ define a parameter range with obvious anomalous magnon softening and damping. The setting of the main parameters is listed in table \ref{settings}.\\
Equations \eqref{Ai_def}, \eqref{results1}, and \eqref{results2}, respectively, predict that the $A_i$ vanish and consequently $\Gamma_\mathbf q\rightarrow0$ for $T\rightarrow0K$ when no magnon-magnon interaction is present. This contradicts the results of other theories of the Kondo lattice model which give finite linewidths at $T=0K$ due to direct contributions to $\Gamma_\mathbf q$ by electron-magnon interactions.\cite{singh,golosov}

\subsection{\label{small_bo}Small band occupation}

\begin{figure}[t]
\includegraphics[width=\mywidth,height=\myheight]{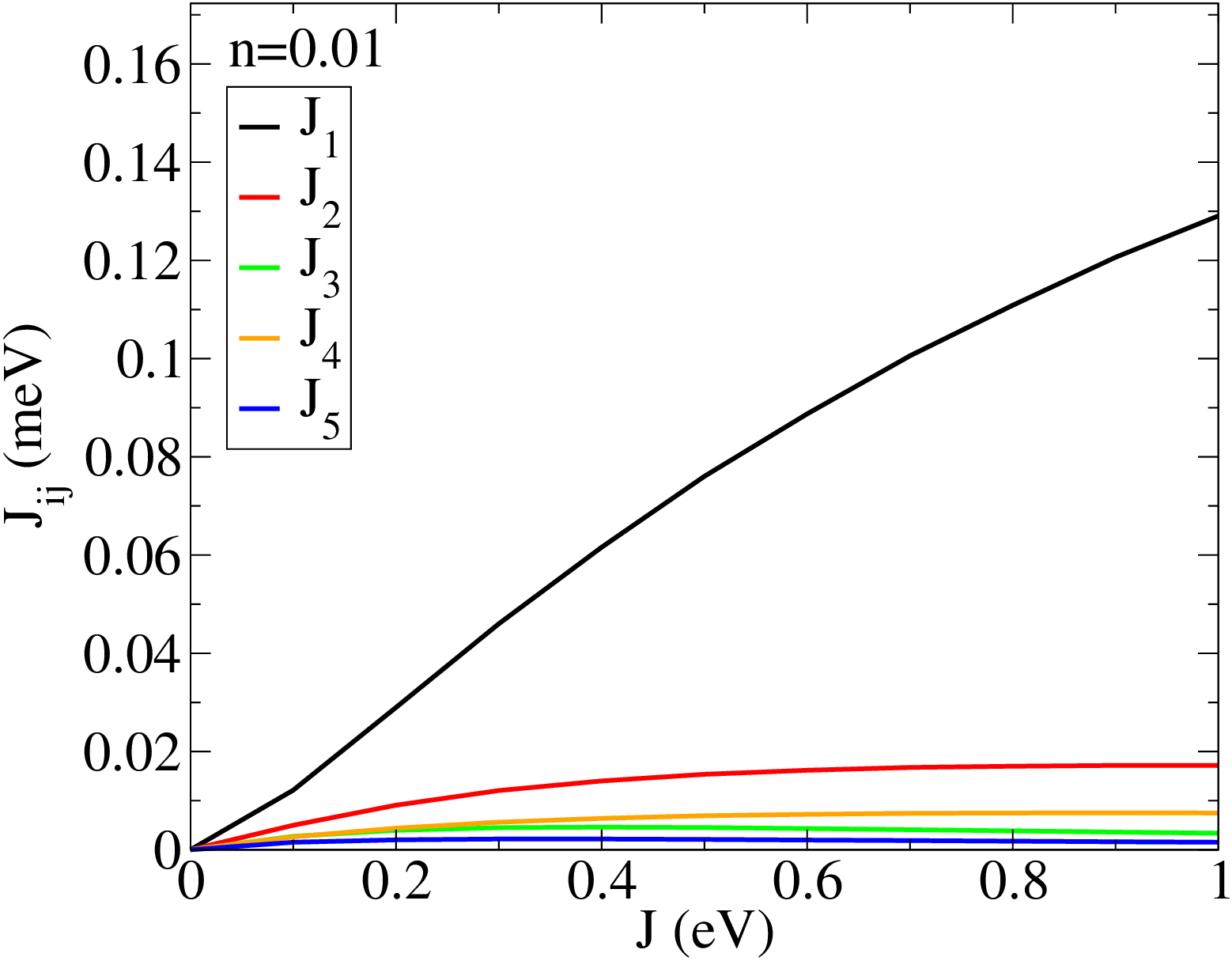}
\caption{(Color online) Small band occupation $n=0.01$. Exchange integrals $J_{ij}$ of the first five shells at $T=0.1K$ as a function of the coupling constant $J$.}
\label{small_bo_Jij}
\end{figure}

\begin{figure}[b]
\includegraphics[width=\mywidth,height=\myheight]{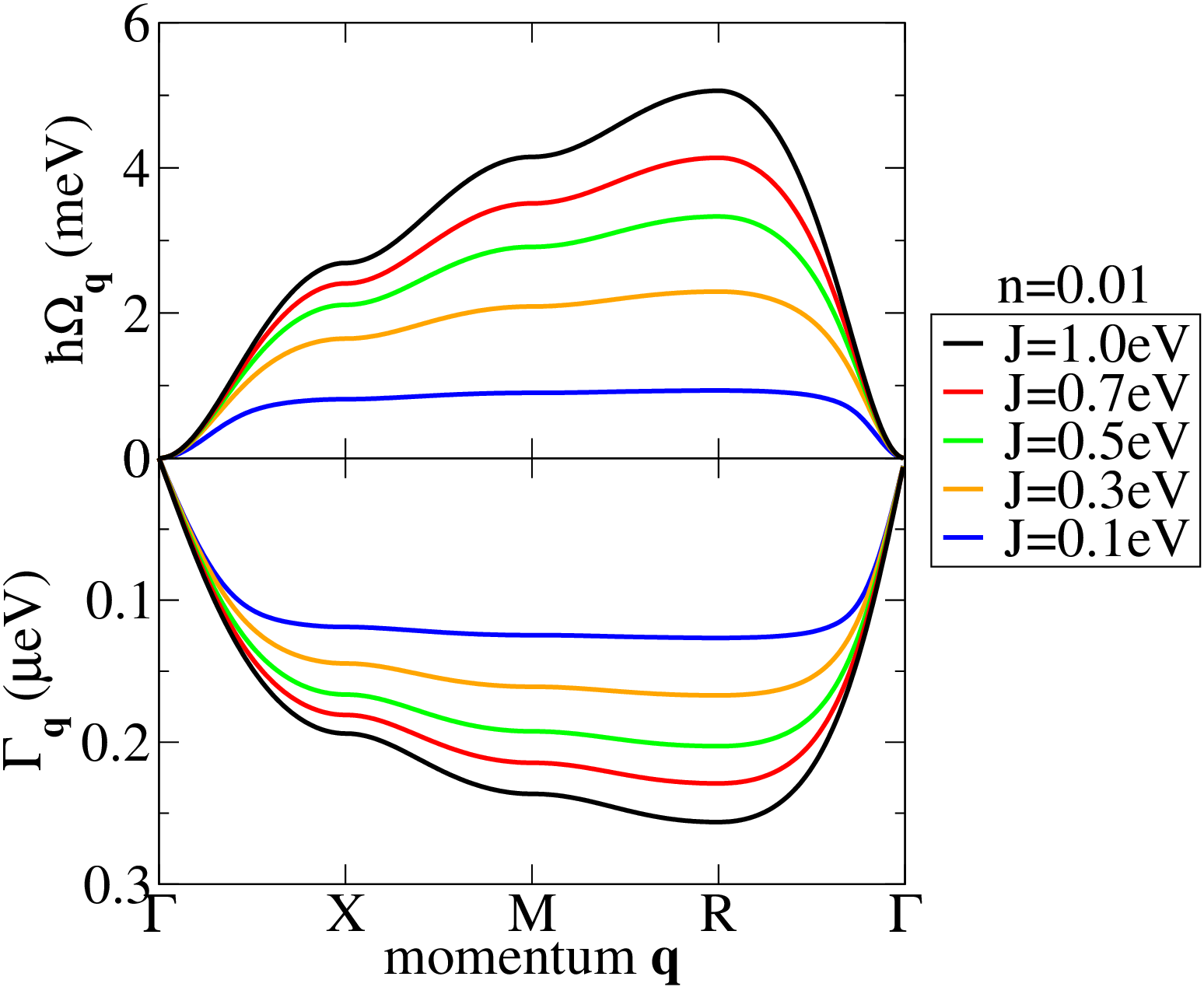}
\caption{(Color online) Small band occupation $n=0.01$. Magnon dispersion relation $\hbar\Omega_\mathbf q$ and momentum dependence of the linewidths $\Gamma_\mathbf q$ for $T=0.1K$ and different values of the coupling constant $J$.}
\label{small_bo_q}
\end{figure}

\begin{figure}[t]
\includegraphics[width=\mywidth,height=\myheight]{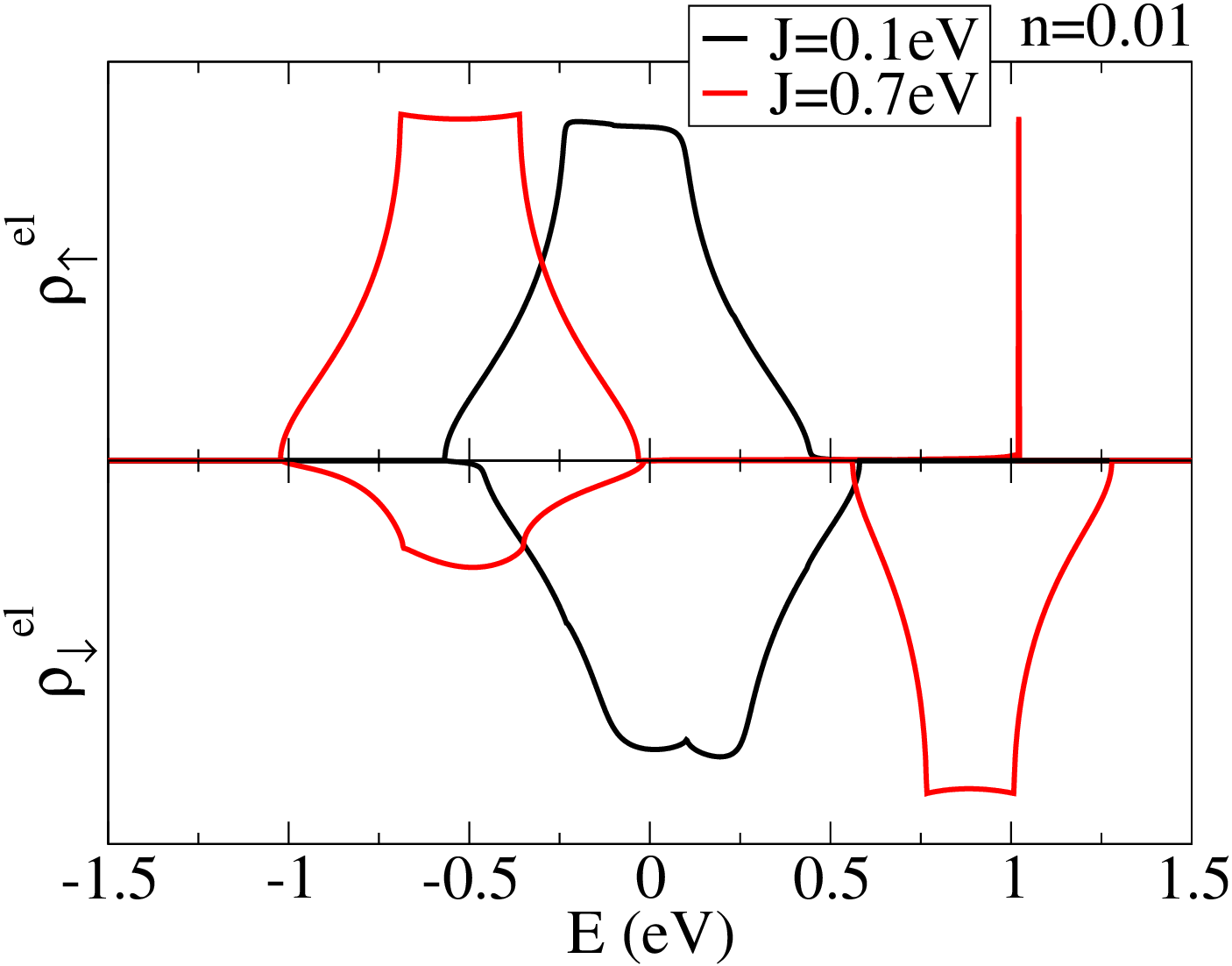}
\caption{(Color online) Small band occupation $n=0.01$. Spin-resolved, local electron density of states $\rho_{\sigma}^\text{el}(E)$ for $T=5K$ and different values of the coupling constant $J$.}
\label{small_bo_el}
\end{figure}

This limiting case is implemented in our calculations by setting $n=0.01$. For all values of $J>0$, the exchange integrals $J_{ij}$ are positive making ferromagnetism possible (Fig. \ref{small_bo_Jij}). The growth of the exchange integrals $J_{ij}$ for increasing $J$ is accompanied by a corresponding growth of both the magnon energies $\hbar\Omega_\mathbf q$ and the linewidths $\Gamma_\mathbf q$ (Fig. \ref{small_bo_q}). For small $J$, we find that $\hbar\Omega_\mathbf q$ and $\Gamma_\mathbf q$ are nearly independent on the momentum $\mathbf q$ since all exchange integrals $J_{ij}$ are of the same order of magnitude. That is why higher exchange integrals  $J_{n>1}$ have great influence on $\hbar\Omega_\mathbf q$ and $\Gamma_\mathbf q$. However, for $J\approx W$, the magnon energy and linewidth are mainly governed by the nearest-neighbour coupling $J_1$, and higher exchange integrals can be neglected. Accordingly, $\hbar\Omega_\mathbf q$ and $\Gamma_\mathbf q$ converge to the common curve shape of a ferromagnetic nearest-neighbours Heisenberg model.\\
The spin-resolved electron density of states (DOS) $\rho^\text{el}_{\sigma}(E)$ (Fig. \ref{small_bo_el}) features the typical properties of the ISA for the case of low temperatures.\cite{nol_isa1} For weak couplings $J\approx0$, there is just a small exchange splitting between $\rho^\text{el}_\uparrow(E)$ and $\rho^\text{el}_\downarrow(E)$. For strong couplings, the $\rho^\text{el}_\downarrow(E)$ band splits into two sub-bands. One is shifted by about $+\frac{J}{2}(S+1)$ to larger energies and is built up by electrons that stabilize their state by permanently absorbing and emitting magnons ("magnetic polaron"). The second band at smaller energies is the scattering band for electrons that have flipped their spin by emitting a magnon. At non-zero temperatures, a high-energy sub-band for $\uparrow$-electrons, too, emerges, mainly provoked by thermally-excited magnons.

\begin{figure}[b]
\includegraphics[width=\mywidth,height=\myheight]{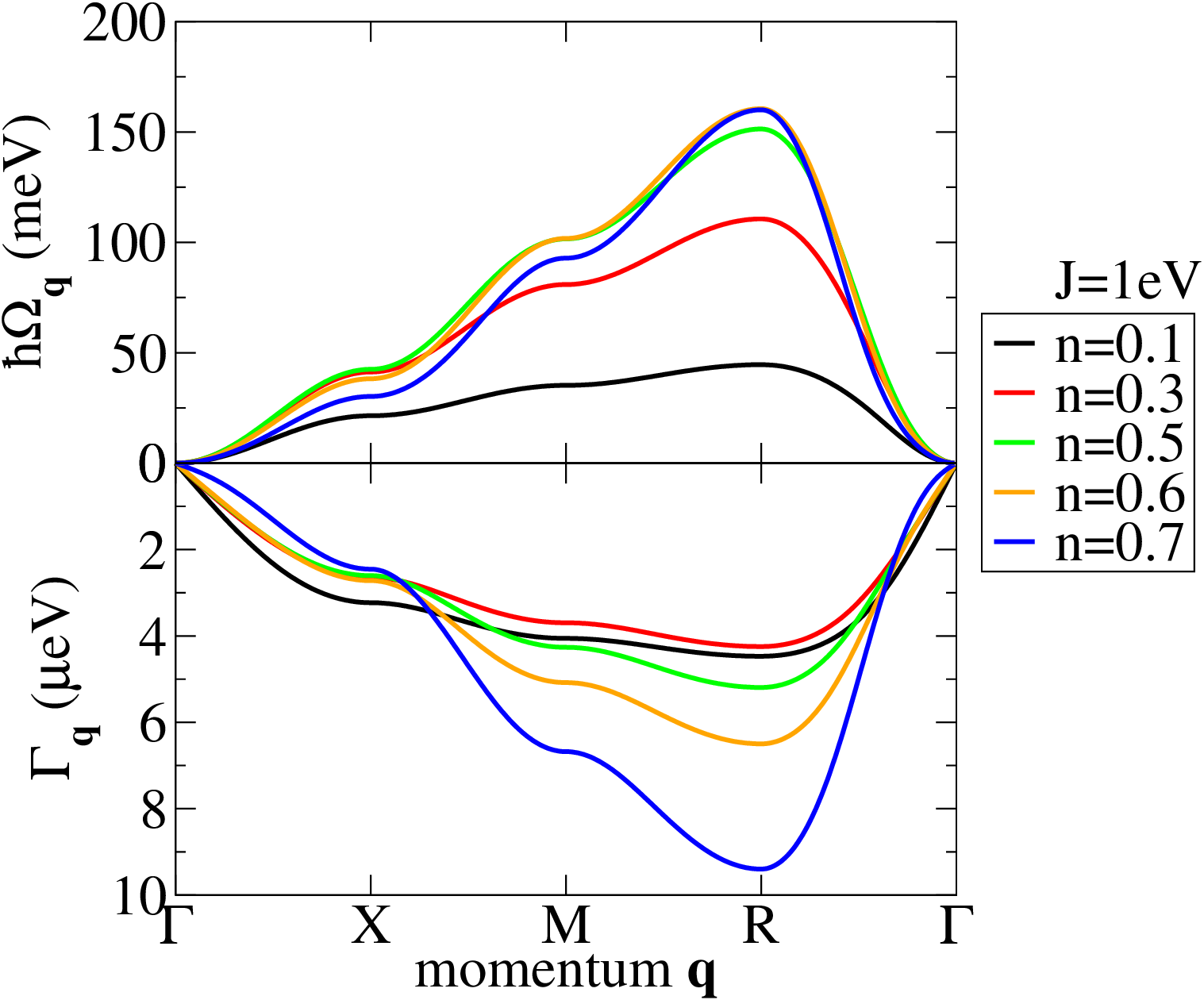}
\caption{(Color online) "Strong coupling" regime $J=1.0eV$. Magnon dispersion relation $\hbar\Omega_\mathbf q$ and momentum dependence of the linewidths $\Gamma_\mathbf q$ for $T=1K$ and different values of the band occupation $n$.}
\label{large_J_q}
\end{figure}

\subsection{\label{large_J}"Strong coupling" regime}

\begin{figure}[t]
\includegraphics[width=\mywidth,height=\myheight]{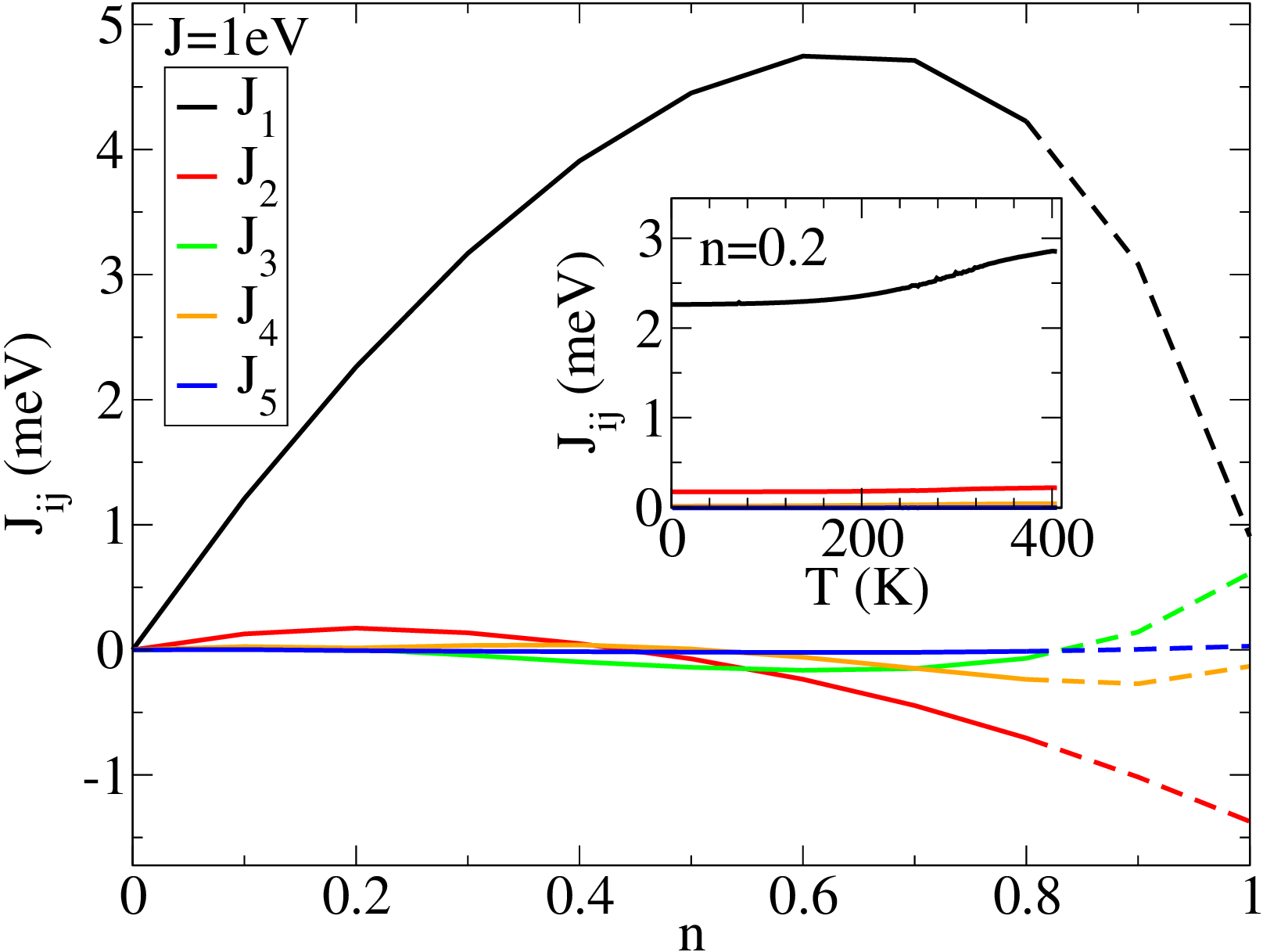}
\caption{(Color online) "Strong coupling" regime $J=1.0eV$. Exchange integrals $J_{ij}$ of the first five shells for $T=1K$ as a function of the band occupation $n$. The broken lines show the behaviour according to Eq. \eqref{self_energy} with the setting $\langle S^z\rangle=S$ for values of $n$ where ferromagnetism is impossible at $T=1K$.\\
The inset shows the temperature dependence of the exchange integrals $J_{ij}(T)$ of the first five shells for $n=0.2$ and temperatures up to $T_\text C$.}
\label{large_J_Jij}
\end{figure}

Although the strong coupling regime is often identified with the condition $J\gg W$, we will use this term for the situation $J=W$ as well, since it marks a threshold in $J$ whereupon no qualitative deviations appear any more in the quantities which we regard here.\\
Starting at small band occupations $n$ and with increasing $n$, the magnon energies $\hbar\Omega_\mathbf q$ grow and the linewidths $\Gamma_\mathbf q$ decline (mainly at the $X$ point, Fig. \ref{large_J_q}). This is made clear by the fact that the exchange integrals $J_{ij}$ first grow because more indirect exchange between the localized spins is possible when there are more conduction band electrons present (Fig. \ref{large_J_Jij}). $\hbar\Omega_\mathbf q$ and $\Gamma_\mathbf q$ reach a maximum (minimum) at about quarter band filling $n\approx0.5$ and take the usual curve shape of a ferromagnetic nearest-neighbours Heisenberg model. When reaching even larger electron densities $n$, this behaviour is reversed: The energies decrease and the linewidths increase (mainly at the $R$ point), relying on negative higher exchange integrals $J_{n>1}<0$ (antiferromagnetic coupling) and the declining nearest-neighbour coupling $J_1$. For band occupations above a critical value of $n\approx0.8$, this trend leads to negative magnon energies that destabilize the ferromagnetic order and prefer antiferromagnetism. The reason why the antiferromagnetic state is more favourable for the case of half band filling $n=1$ can be understood with Pauli's exclusion principle: The electrons can reduce their energy when they virtually hop to adjacent lattice sites, which is only possible if there is no electron with the same spin present.\\
A study of the temperature dependence (inset of Fig. \ref{large_J_Jij}) reveals that primarily the nearest-neighbour coupling $J_1$ is enhanced for rising temperatures, while higher exchange integrals remain temperature independent.

\subsection{Anomalous Magnon Softening and Damping}

\begin{figure}[t]
\includegraphics[width=\mywidth,height=\myheight]{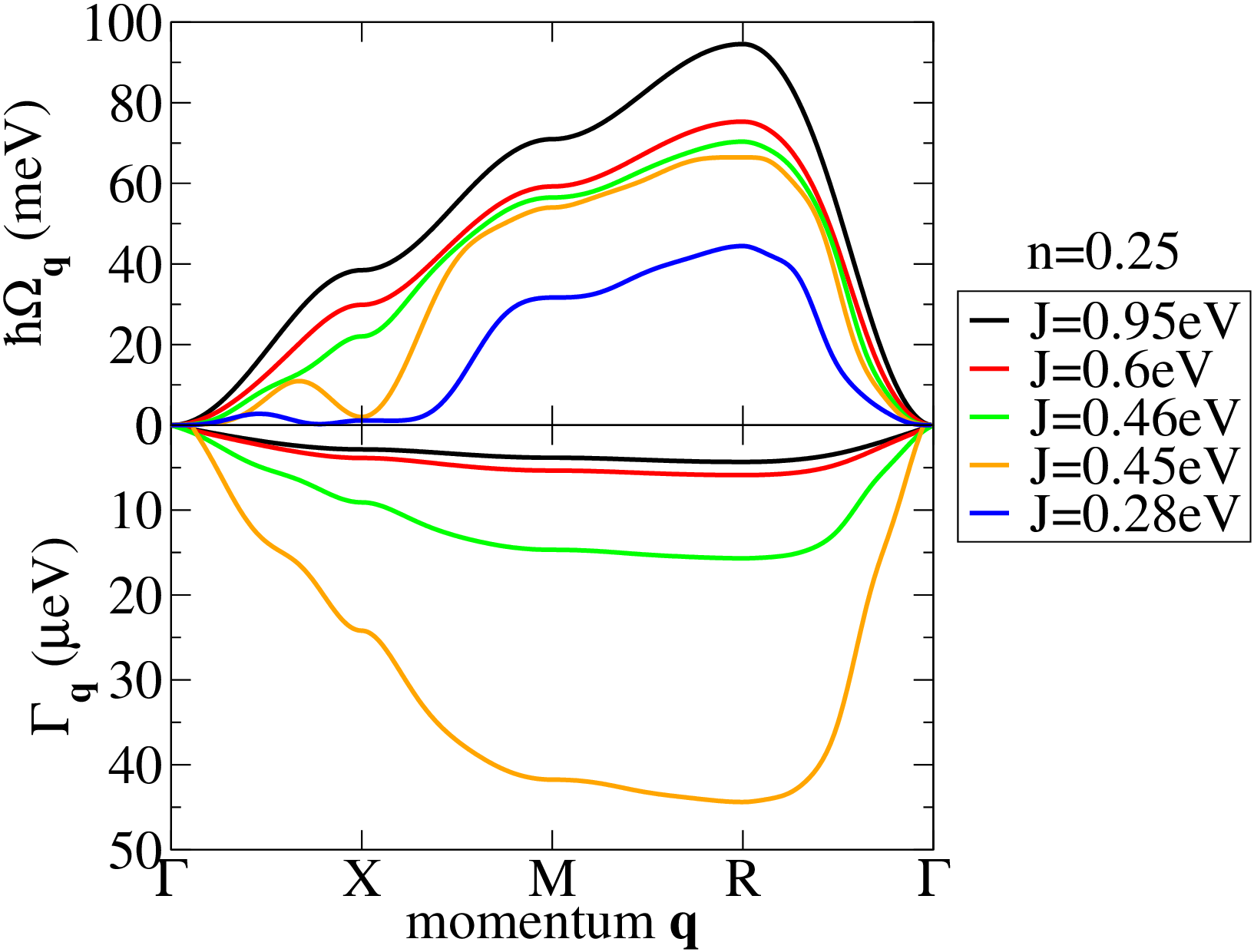}
\caption{(Color online) Band occupation $n=0.25$. Magnon dispersion relation $\hbar\Omega_\mathbf q$ and momentum dependence of the linewidths $\Gamma_\mathbf q$ for $T=1K$ and different values of the coupling constant $J$. For couplings below $J=0.45eV$, the equation for $\Gamma_\mathbf q$ (Eq. \eqref{results2}) yields no real solution at $T=1K$, indicating strong antiferromagnetic tendencies in the exchange integrals $J_{ij}$.}
\label{anom_q}
\end{figure}

\begin{figure}[b]
\includegraphics[width=\mywidth,height=\myheight]{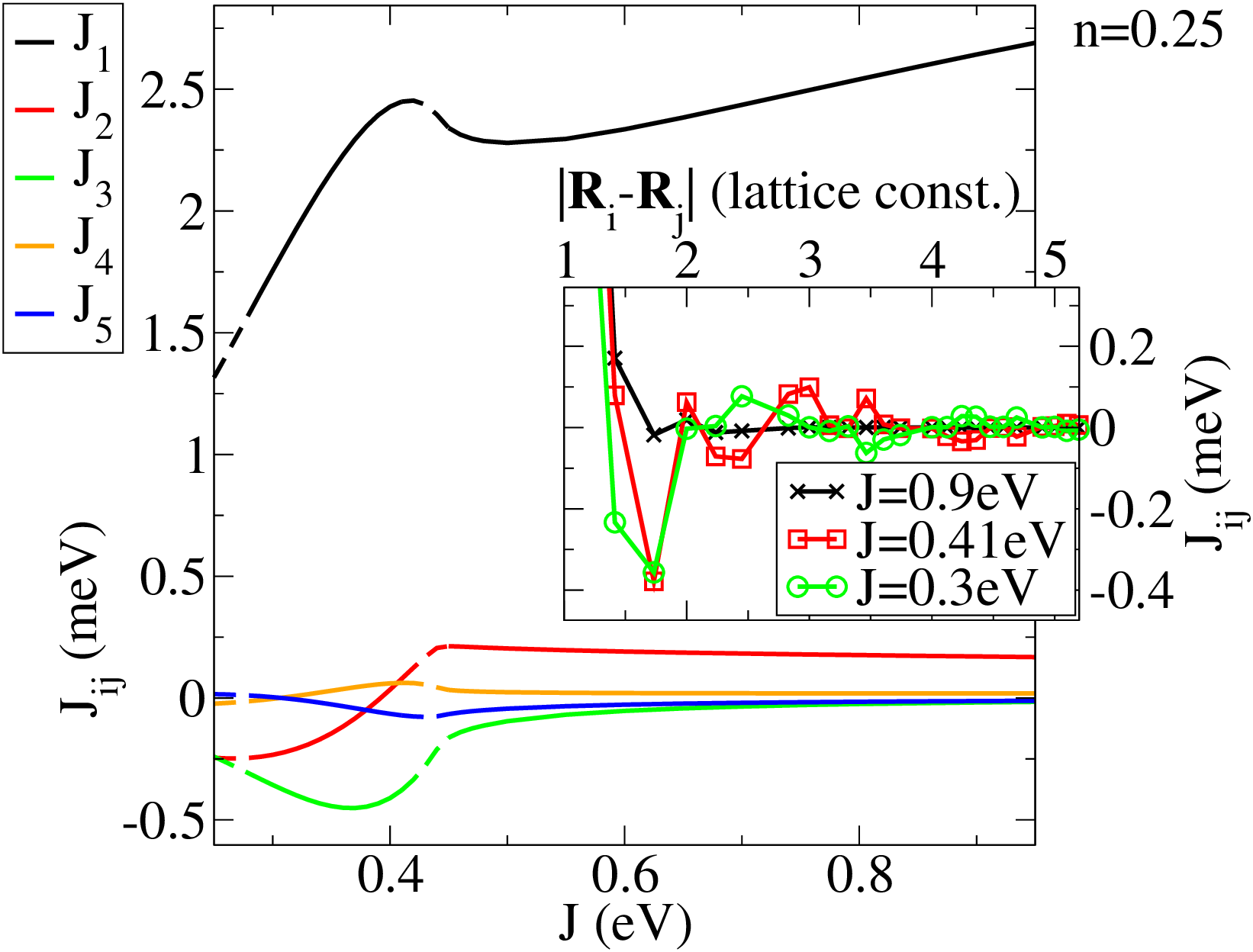}
\caption{(Color online) Band occupation $n=0.25$. Exchange integrals $J_{ij}$ of the first five shells for $T=1K$ as a function of the coupling constant $J$. The broken lines show the behaviour according to Eq. \eqref{self_energy} with the setting $\langle S^z\rangle=S$ for values of $J$ where ferromagnetism is impossible at $T=1K$.\\
The inset shows the dependence on the distance $|\mathbf R_i-\mathbf R_j|$ of the exchange integrals $J_{ij}$ for $T=1K$ and different values of $J$. The lines are a guide for the eyes.}
\label{anom_Jij}
\end{figure}

\begin{figure}[t]
\includegraphics[width=\mywidth,height=\myheight]{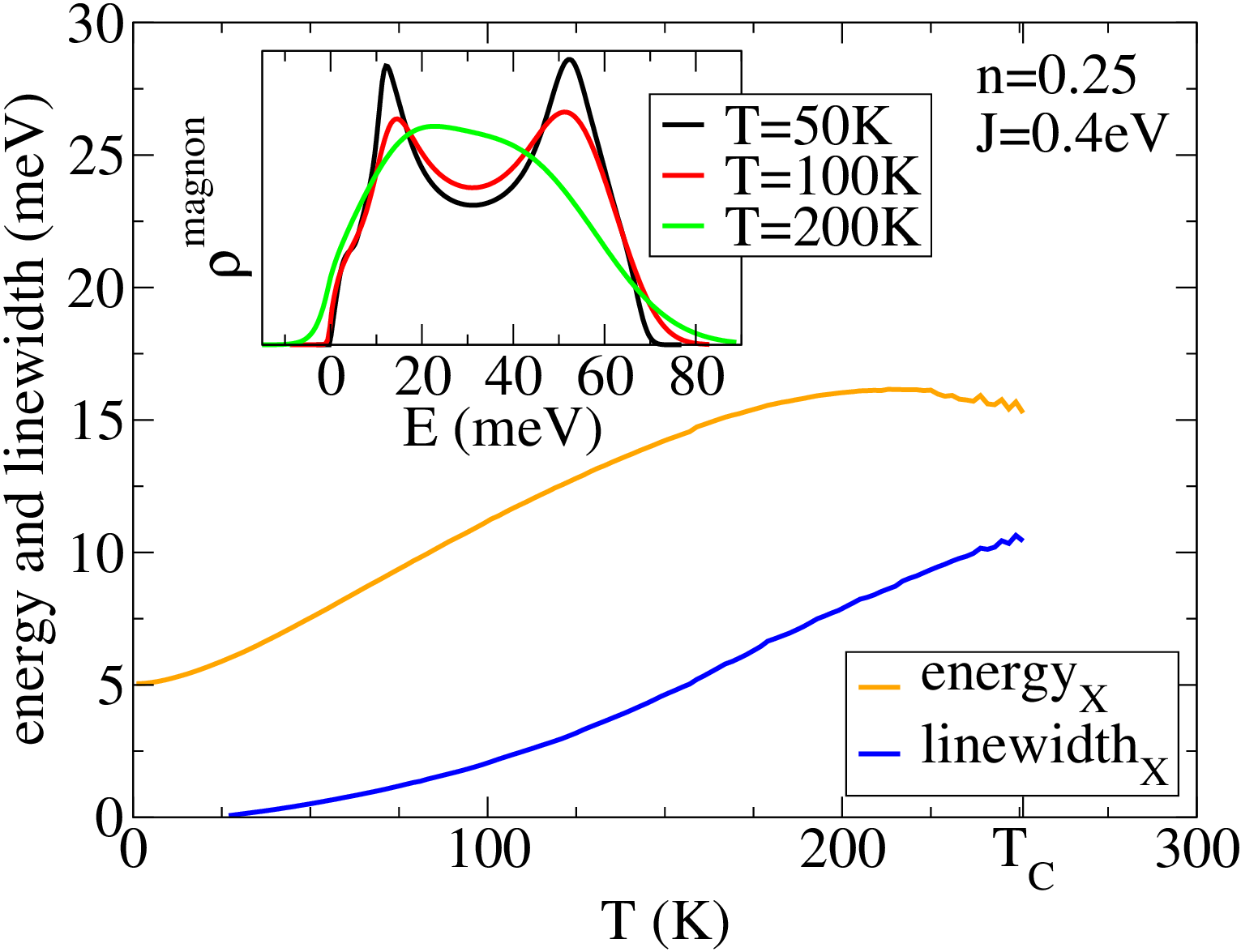}
\caption{(Color online) Band occupation $n=0.25$. Temperature dependence of the magnon energy $\hbar\Omega_X(T)$ and the linewidths $\Gamma_X(T)$ at the $X$ point for $J=0.4eV$. For temperatures below $T=27K$, the equation for $\Gamma_\mathbf q$ (Eq. \eqref{results2}) yields no real solution, indicating strong antiferromagnetic tendencies in the exchange integrals $J_{ij}$.\\
The inset shows the magnon density of states $\rho^\text{magnon}(E)$ for $J=0.4eV$ at different temperatures up to $T_\text C$.}
\label{anom_T}
\end{figure}

In the case of intermediate couplings $J$ and intermediate band occupations $n$, the magnon energies and linewidths sensitively depend on both $J$ and $n$. We investigate the situation for different values of $J$ in the vicinity of a band filling of $n=0.25$ and $J=0.4eV$.\\
Higher exchange integrals $J_{n>1}$ are comparatively large and often negative for $J\lesssim0.5eV$ (Fig. \ref{anom_Jij}), giving rise to distinct deformations of the magnon dispersion relation $\hbar\Omega_\mathbf q$ and of the curve shape of $\Gamma_\mathbf q$ mainly at the Brillouin zone boundaries (Fig. \ref{anom_q}). The strongest modifications of the magnon dispersion relation occur around the $X$ point along with smaller ones at the $R$ point. Below a critical $J=0.28eV$, parts of the Brillouin zone evolve where the magnon energy becomes negative and for this reason ferromagnetism unstable. The linewidths exhibit deviations from the common behaviour of a ferromagnetic nearest-neighbours Heisenberg model between $\Gamma$ and $X$ point and at the $M$ point. Compared to $X$ and $M$ point, they are unusually small at the $R$ point, which leads to unusually long magnon lifetimes. When approaching the critical $J$, the linewidths dramatically increase as expected from a Heisenberg model near the transition from the ferromagnetic to the paramagnetic state.\\
Furthermore, we find distinct long-range oscillations of the exchange integrals between ferromagnetic $J_{ij}>0$ and antiferromagnetic coupling $J_{ij}<0$ qualitatively similiar to the conventional RKKY theory (inset of Fig. \ref{anom_Jij}).\\
It should be pointed out that all the mentioned effects are consequences of solely electron-magnon and magnon-magnon interactions.\\
When we increase the temperature (Fig. \ref{anom_T}), $\hbar\Omega_\mathbf q$ and $\Gamma_\mathbf q$ reveal unexpected characteristics. The deviations at the $X$ point in relation to the usual Heisenberg model are reduced, and the magnon energies grow with increasing $T$ - even at temperatures near $T_\text C$. This behaviour radically differs from the usual behaviour observed for a Heisenberg model (e.g. Fig. \ref{comparison}). It relies on the growth of the corresponding exchange integrals $J_{ij}(T)$, mainly of $J_1$ and $J_4$, favouring the ferromagnetic order. Furthermore and in contrast to the Secs. \ref{small_bo} and \ref{large_J} and to the results for EuO in Sec. \ref{compare_EuO}, we observe a relatively strong temperature dependence of the linewidths, which behave like $\Gamma_X(T)\sim T^{2}$ for $J=0.4eV$ (Fig. \ref{anom_T}).\\
Although the magnon density of states (inset of Fig. \ref{anom_T}) contains the characteristic tight-binding curve shape owing to a dominating nearest-neighbour exchange $J_1$, it exhibits deviations at energies $E\approx30meV$ and low temperatures due to the deformations in $\hbar\Omega_\mathbf q$ and $\Gamma_\mathbf q$. The rise in temperature firstly leads to larger spectral weight at $E\approx30meV$ and finally washes out the structure because of the larger linewidths near $T_\text C$.\\
Anomalous magnon softening and damping can be detected in neutron scattering experiments with manganites.\cite{zhang_review,dai,ye,ye2,moussa} From the theoretical point of view, anomalous softening at the $X$ point has been reported by other authors \cite{santos, vogt} confirming the parameter range of intermediate $J$ and $n$. A theory that features anomalous magnon damping has been proposed by Pandey et al.\cite{singh} Therein, the spin operators $\mathbf S_i$ in the Hamiltonian of the Kondo lattice model \eqref{kondo_hamiltonian} are fermionized by localized electrons in atom orbitals. An "inverse-degeneracy expansion approach" is applied, describing the diagrammatic contributions in powers of the inverse number of orbitals incorporated in the calculations. The authors have mentioned clear differences from the common behaviour of a Heisenberg model for $\hbar\Omega_\mathbf q$ at $X$ and $M$ point and for $\Gamma_\mathbf q$ between $M$ and $R$ point, even for large $J=6W$ and $J\rightarrow\infty$. In the double exchange limit $J\rightarrow\infty$, other authors\cite{golosov} have found anomalous softening and damping, too.

\section{Conclusions}

We have presented an approach for calculating the magnon energies $\hbar\Omega_\mathbf q$ and linewidths $\Gamma_\mathbf q$ for the Kondo lattice model and examined their dependencies on the band occupation $n$, the coupling constant $J$, and the temperature $T$. Likewise, our ansatz allows us to study other interesting quantities such as the electron and magnon density of states $\rho(E)$ or the exchange integrals $J_{ij}$. We have studied it for small band occupation, the case of $J=W$, and for intermediate $J$ and $n$ where the magnon spectrum shows anomalies at the Brillouin zone boundaries. These deviations can be explained by partial antiferromagnetic indirect exchange between the localized spins as a consequence of electron-magnon and magnon-magnon interaction. We have demonstrated that the deformations of the magnon dispersion relation due to anomalous softening become smaller as the temperature rises. As mentioned above, these anomalies are caused by electron-magnon and magnon-magnon interactions only. This finding could permit a better understanding of the origin of similar anomalies in real materials.\\
Note that our method is also applicable directly to a pure Heisenberg model with given exchange integrals $J_{ij}$.\\
\\
When comparing our numerical results with experimental data of La$_{0.7}$Ca$_{0.3}$MnO$_3$,\cite{dai,ye,moussa} we state differences in $\hbar\Omega_\mathbf q$ and $\Gamma_\mathbf q$. Although it is difficult to compare theory and experiment without knowledge of the electronic band structure, it can be argued that the differences can be ascribed to the Hamiltonian \eqref{kondo_hamiltonian} which we have used to derive our results \eqref{results1} and \eqref{results2}. Namely, it does not involve electron-electron, spin-spin, or electron-phonon interactions, even though they are regarded as essential for the manganites. The linewidths calculated by our method turn out to be too small,\cite{dai, golosov} which can be explained by the absent electron-phonon interaction. Moreover, it has been shown that the incorporation of an on-site interaction $U$ between the itinerant electrons changes the magnon dispersion relation drastically.\cite{kapetanakis} In order to take these terms into account, our approach can be extended.\cite{nol_isa2, stier1}

\begin{acknowledgments}
This work was supported by the SFB 668 of the "Deutsche Forschungsgesellschaft".
\end{acknowledgments}


\end{document}